\documentclass[a4paper,12pt]{article}
\usepackage{amsmath,amssymb,amsfonts,amsthm}

\newcommand{\dv}{\, d\mu_0}

\newcommand{\dvh}{\, d\mu_h}

\newcommand{\ds}{\, ds}

\newcommand{\dsh}{\, ds_h}

\newcommand{\dshp}{\, ds_{\tilde{h}}}

\newcommand{\Rt}{\mathbb{R}^3}

\newcommand{\mf}{\mathcal{M}}

\newcommand{\vv}{\alpha}

\newcommand{\vs}{\gamma}

\newcommand{\vY}{y}

\newcommand{\af}{\mathcal{A}}

\title{A variational principle for  stationary, axisymmetric
  solutions of Einstein's  equations}
\author{Sergio Dain\\
  Max-Planck-Institut f\"ur Gravitationsphysik\\
  Am M\"uhlenberg 1\\
  14476 Golm\\
  Germany}

\begin{document}
\maketitle

\begin{abstract}
  Stationary, axisymmetric, vacuum, solutions of Einstein's equations
  are obtained as critical points of the total mass among all
  axisymmetric and $(t,\phi)$ symmetric initial data with fixed
  angular momentum. In this variational principle the mass is written
  as a positive definite integral over a spacelike hypersurface. It is
  also proved that if absolute minimum exists then it is equal to the
  absolute minimum of the mass among all maximal, axisymmetric,
  vacuum, initial data with fixed angular momentum.  Arguments are
  given to support the conjecture that this minimum exists and is the
  extreme Kerr initial data.
\end{abstract}

\section{Introduction}
In an axisymmetric, vacuum, gravitational collapse the total angular momentum
is a conserved quantity. Therefore, if we assume, according to the
standard picture of the gravitational collapse, that the final state will
be a Kerr black hole the following inequality should hold for every
axisymmetric, vacuum, asymptotically flat, complete, initial data set
\begin{equation}
  \label{eq:14}
\sqrt{|J|}\leq m,
\end{equation}
where $m$ is the mass of the data and $J$ the angular momentum in the
asymptotic region.  Moreover, the equality in \eqref{eq:14} should
imply that the data is an slice of the extreme Kerr black hole. A
counter example to \eqref{eq:14} will provide a regular vacuum data
that do not settle down to a Kerr black hole. For a more detailed
discussion of the motivations and relevance of \eqref{eq:14} and
related inequalities see \cite{Friedman82}, \cite{Horowitz84}  and
\cite{Dain02c}.

Inequality \eqref{eq:14} is a property of the spacetime and not only
of the data, since both quantities $J$ and $m$ are independent of the
slicing. It is in fact a property of axisymmetric, vacuum, black
holes spacetimes, because a non zero $J$ (in vacuum) implies a non
trivial topology on the data and this is expected to signal the
presence of a black hole. Note, however, that the mass in
\eqref{eq:14} is a global quantity but the angular momentum is a
quasilocal quantity because we have assumed axial symmetry. Without
axial symmetry we still have $J$ defined as a global quantity at
spacelike infinity, but \eqref{eq:14} is not longer true in this case.
A more subtle question is whether \eqref{eq:14} is true where both $m$
and $J$ are quasilocal quantities, that is, whether \eqref{eq:14} is
in fact a quasilocal property of the black hole. In general there is
no unique definition of quasilocal mass (see the recent review on the
subject \cite{Szabados04}). However, a remarkable counter example was
found in \cite{Ansorg05} in which there is a clear quasilocal mass
definition (the Komar mass) and inequality \eqref{eq:14} is violated
at the quasilocal level.  Finally, let us note that \eqref{eq:14} is
false for black holes in higher dimensions (see, for example,
\cite{Horowitz05} and reference therein). 

The inequality \eqref{eq:14} suggests the following variational
principle: 
\begin{itemize}
\item[(i)] \emph{The extreme Kerr initial data is the absolute minimum
    of the mass among all  axisymmetric, vacuum, asymptotically
    flat and complete initial data with fixed angular momentum.}
\end{itemize}
So far, there is no proof of \eqref{eq:14}. A promising strategy to
prove it is to use the variational formulation (i).  
In this article we will prove the following results, which are a step
forward in this direction.

The first result is the following  related variational principle:
\begin{itemize}
\item[(ii)] 
  \emph{The critical points of the mass among all the axisymmetric,
    $(t,\phi)$ symmetric, asymptotically flat data are the stationary,
    axisymmetric solutions.}
\end{itemize}
A spacetime is defined to be $(t,\phi)$ symmetric if it is symmetric
under a simultaneous change of sign of the time coordinate $t$ and the
axial angle $\phi$. A data is called $(t,\phi)$ symmetric if its
evolution is a $(t,\phi)$ symmetric spacetime. These data are also
known as ``momentarily stationary data'' (see \cite{Bardeen70} for
more details).  The variational principle (ii) was proved by Bardeen 
\cite{Bardeen70}, who also included matter in the formulation. It was
also studied by Hawking  \cite{Hawking73} for black holes including boundary
terms.  However, in all these works the mass is not written as a
positive definitive integral (see the discussion of section VIII in
\cite{Bardeen70}). Therefore, it is not possible to relate (ii) with
(i) in these formulations. 
In this article we will prove (ii)
using the mass formula discovered by Brill \cite{Brill59}, which is
a positive definitive integral over the slice.  
Using this formulation 
of (ii) we will be able to prove the following: 

\begin{itemize}
\item[(i')] \emph{If the absolute minimum of the mass among all
    axisymmetric, $(t,\phi)$ symmetric, vacuum, asymptotically flat
    and complete initial data with fixed angular momentum exists, then
    it is equal to the absolute minimum of the mass among all maximal,
    axisymmetric, vacuum, asymptotically flat and complete initial
    data with fixed angular momentum. Moreover, the absolute minimum
    is stationary.}
\end{itemize}
That is, we have essentially reduced the variational problem (i) to
the $(t,\phi)$ symmetric case.  Note that we have included in (i') the
condition that the data are maximal (i.e. the trace of the second
fundamental form is zero).  This is a technical assumptions which
simplifies considerably the analysis, but the statement is expected to
be valid without it.

There exist other variational formulations of the stationary,
axisymmetric,  equations, see \cite{Matzner67}
\cite{Neugebauer92}. Particularly interesting in the present context
is the variational formulation given by Carter \cite{Carter73} which
is based in the Ernst formulation \cite{Ernst68}. 
There exist a remarkable connection between (ii) in the present
formulation and Carter's variational principle,  we will prove that the
Lagrangians differ only by a (singular) boundary term.

\section{Axially symmetric initial data and Brill proof of the
  positive mass theorem}\label{sec:axially-symm-init}

In this section we review Brill's positive mass theorem for
axisymmetric data \cite{Brill59}. The original proof was for
time-symmetric data in $\Rt$, here we slightly extend it to include
maximal data and non-trivial topologies.
  
An initial data set for Einstein's vacuum equations consists in a
3-manifold $S$, a Riemannian metric $\tilde{h}_{ab}$, and a symmetric
tensor field $\tilde{K}^{ab}$ such that the vacuum constraint
equations
\begin{align}
 \label{const1}
 \tilde D^b \tilde K_{ab} -\tilde D_a \tilde K= 0,\\
 \label{const2}
 \tilde R + \tilde K^2-\tilde K_{ab} \tilde K^{ab}=0,
\end{align}
are satisfied on $S$; where $\tilde{D}_a$ and $\tilde R$ are the
Levi-Civita connection and the Ricci scalar associated with
$\tilde{h}_{ab}$, $\tilde K=\tilde h^{ab} \tilde K_{ab}$, and the
indexes are moved with the metric $\tilde h_{ab}$ and its inverse
$\tilde h^{ab}$.

We will assume that the initial data are axially symmetric, that is,
there exist an axial Killing vector $\eta^a$ such that
 \begin{equation}
  \label{eq:8}
 \pounds_\eta \tilde h_{ab}=0, \quad  \pounds_\eta \tilde K_{ab}=0,
\end{equation}
where $\pounds$ denotes the Lie derivative.  The Cauchy development
of such initial data will be an axially symmetric spacetime.

The Killing vector $\eta^a$ is assumed to be orthogonal with respect
to $\tilde h_{ab}$ to a family of 2-surfaces in $S$.  Under these
conditions, the metric $\tilde h_{ab}$ can be characterized by two
functions, one is essentially the norm of the Killing vector and the
other is a conformal factor on the 2-surfaces. We make explicit the
choice of the free functions as follows.  Let $(\rho,z,\phi)$ be local
coordinates in $S$ such that the metric has following form
\begin{equation}
  \label{eq:1}
  \tilde h_{ab}=\psi^4h_{ab},
\end{equation}
where the conformal metric $h_{ab}$ is given
\begin{equation}
  \label{eq:105}
  h= e^{-2q}(d\rho^2+dz^2)+\rho^2d\varphi^2,
\end{equation}
and $q, \psi$ are functions which depend only on $z$ and $\rho$ with
$\psi>0$.  The vector $\eta^a= (\partial/\partial \varphi)^a$ is a
Killing vector of both metrics $\tilde h_{ab}$ and $h_{ab}$. The norm
of $\eta^a$ with respect to the physical metric will be denoted by
$X$, (i.e. $X =\eta^a\eta^b\tilde h_{ab}=\psi^4\rho^2$), the norm of
$\eta^a$ with respect to the conformal metric is given by $\rho^2=
\eta^a\eta^b h_{ab}$.

We define the following quantity
\begin{equation}
  \label{eq:52}
  J(\Sigma)= \oint_{\Sigma} \tilde \pi_{ab}\eta^a \tilde n^b \dshp,
\end{equation}
where $\tilde \pi_{ab}= \tilde K_{ab}-\tilde h_{ab}\tilde K$, $\Sigma$
is any closed 2-surface, $\tilde n^a$ is the unit normal vector to
$\Sigma$ with respect to $\tilde h_{ab}$ and $\dshp$ is the area
element of $\Sigma$ with respect to $\tilde h_{ab}$.  Equation
\eqref{const1} and the Killing equation imply that the vector $\tilde
\pi_{ab}\eta^a$ is divergence free. If $\Sigma$ is the boundary of
some compact domain $\Omega\subset S$, by the Gauss theorem, we have
$J(\Sigma)=0$.  For example, if $S=\Rt$ then $J(\Sigma)=0$ for all
$\Sigma$.  In an asymptotically flat data, $J(\Sigma_{\infty})$ gives
the total angular momentum, where $\Sigma_{\infty}$ is any closed
surface in the asymptotic region. Then, the angular momentum will be
zero unless $\Sigma_{\infty}$ is not the boundary of some compact domain
contained in $S$. 

In order to have non zero angular momentum we will allow $S$ to have
many asymptotic ends\footnote{There is an interesting alternative (not
  included here) discussed in \cite{Friedman82} to allow non zero
  angular momentum: the interior of the manifold is assumed to be
  compact and non simply connected with a pseudo axial Killing
  vector.}.  Let $i_k$ a finite number of points in $\Rt$.  The
manifold $S$ is assumed to be $\Rt\setminus \sum_k i_k$. The points
$i_k$ will represent the extra asymptotic ends, at those points we
will impose singular boundary conditions for $\psi$. The be consistent
with the axial symmetry assumption the points $i_k$ should be located
on the axis $\rho=0$.
 
In addition to axial symmetry we will assume that the data are maximal
\begin{equation}
  \label{eq:3}
  \tilde K =0.
\end{equation}
By equation \eqref{const2} this implies that $\tilde R$ is positive,
this will be essential in order to extend Brill's proof to non-time
symmetric data.

Define the conformal second fundamental form by ${K}^{ab} =
\psi^{10}\tilde K^{ab}$. Using \eqref{eq:3} and \eqref{eq:8} we obtain
\begin{equation}
  \label{eq:12}
\pounds_\eta K_{ab}=0, \quad    K =0.
\end{equation}

The constraint equations \eqref{const1}--\eqref{const2} can be
written as equations for $K_{ab}$ and $\psi$ using the well known
conformal method (see, for example, \cite{Bartnik04b} and reference
therein)
\begin{align}  
\label{diver}
D_a K^{ab} &= 0,\\
\label{Lich}
D^aD_a \psi-\frac{1}{8}R \psi &=-\frac{1}{8}K_{ab}K^{ab}\psi^{-7},
\end{align}
where $D_a$ and $R$ are the Levi-Civita connection and the Ricci
scalar associated with the conformal metric $h_{ab}$. In these
equations, the indexes are moved with the conformal metric $h_{ab}$
and its inverse $ h^{ab}$.

The function $q$ is assumed to be smooth with respect to the
coordinates $(\rho,z)$. At the axis we impose the regularity
condition
\begin{equation}
  \label{eq:9}
  q(\rho=0,z)=0.
\end{equation}
Note that condition \eqref{eq:9} includes the points $i_k$. These
points are assumed to be regular points of the conformal metric
$h_{ab}$, that is, $h_{ab}$ is well defined in $\Rt$.

We assume the following fall-off condition at infinity
\begin{equation}
  \label{eq:10}
  q=o(r^{-1}), \quad q_{,r}=o(r^{-2}),
\end{equation}
where $r=\sqrt{\rho^2+z^2}$ and a comma denotes partial derivatives.
This fall off conditions imply that the total mass of the conformal
metric $h_{ab}$ is zero.

At infinity, the conformal factor $\psi$ and the conformal second
fundamental form satisfy 
\begin{equation}
  \label{eq:6}
\psi-1 =O(r^{-1}), \quad  \psi_{,r} = O(r^{-2}), 
\end{equation}
and
\begin{equation}
  \label{eq:39}
  K_{ab}=O(r^{-2}). 
\end{equation}
Under these assumptions the total mass of the physical metric is given
by
\begin{equation}
  \label{eq:111}
  m=\frac{-1}{2\pi}\lim_{r\rightarrow \infty } \oint_{\Sigma_r} n^a D_a \psi
  \dsh,
\end{equation}
where $\Sigma_r$ are the 2-spheres $r=constant$, $n^a$ is the unit
normal, with respect to $h_{ab}$, pointed outwards and $\dsh$ is the
area element of $\Sigma$ with respect to $h_{ab}$.

The Ricci scalar $R$ of the conformal metric \eqref{eq:105} is given
by
\begin{equation}
  \label{eq:106}
  R=2e^{2q}(q_{,\rho \rho }+q_{,z z }).
\end{equation}
We have the important equation
\begin{equation}
  \label{eq:109}
  \int_{\Rt} R \dvh=0,
\end{equation}
where $\dvh$ is the volume element of the metric $h_{ab}$ To prove
this, note that $\dvh=\rho e^{-2q}d\rho dz d\phi$, then
\begin{align}
  \label{eq:110}
  \int_{\Rt} R \dvh &= 4\pi \int_0^\infty d \rho \int_{-\infty}^\infty
  d z \, (q_{,\rho \rho }+q_{,z z})  \rho \\
  & = 4\pi \int_0^\infty d \rho \int_{-\infty}^\infty d z \, \left(
    (\rho q_{,\rho} - q)_{,\rho }+ (\rho q_{,z})_{,z} \right),
\end{align}
we use the divergence theorem in two dimension to transform this
volume integral in a boundary integral over the axis $\rho=0$ and
infinity. The boundary integral at the axis vanishes since $q$
satisfies \eqref{eq:9} and at infinity it also vanishes because of
\eqref{eq:10}.

Since $\lim_{r\rightarrow \infty }\psi =1$, we have an equivalent
expression for the mass
\begin{equation}
  \label{eq:111b}
  m=\frac{-1}{2\pi}\lim_{r\rightarrow \infty } \oint_{\Sigma_r} \frac{
  n^aD_a \psi}{\psi} \dsh.
\end{equation}
We use the identity
 \begin{equation}
   \label{eq:112}
   D_a\left ( \frac{D^a\psi}{\psi}\right)= \frac{D^aD_a \psi}{\psi}-
   \frac{D_a \psi D^a\psi}{\psi^2},
 \end{equation}
 the constraint equation \eqref{Lich}, equation (\ref{eq:109}) and the
 mass formula \eqref{eq:111b} to obtain the final expression
\begin{equation}
  \label{eq:114}
  m= \frac{1}{2\pi}\int_{\mathbb{R}^3}\left(\frac{K^{ab}K_{ab}}{8\psi^8}+
  \frac{D_a\psi D^a\psi}{\psi^2}
\right) \dvh, 
\end{equation}
which is definite positive. To obtain \eqref{eq:114} from
\eqref{eq:112} we have assumed that the boundary integral around the
singular points $i_k$ vanishes, that is
\begin{equation}
  \label{eq:23}
 \lim_{r_k\rightarrow 0 } \oint_{\Sigma_{r_k}} \frac{
  n^aD_a \psi}{\psi} \dsh=0,
\end{equation}
where $r_k$ is the distance to the point $i_k$. This condition (which
is, of course, trivially satisfied when the topology of the physical
data is $\Rt$) allows for a singular behavior of $\psi$ at $i_k$
which in particular include the case where $i_k$ are asymptotically
flat ends. Near an asymptotically flat end $i_k$ the conformal factor
satisfies $\psi=O(r_k^{-1})$, $\psi_{,r}=O(r_k^{-2})$ which imply
\eqref{eq:23}.  To illustrate this, consider the following two
examples.

The Schwarzschild initial data in isotropic coordinates is
time-symmetric ($K_{ab}=0$) and conformally flat ($q=0$). In this case
we have one point $i_0$ located at the origin and the conformal
factor is given by
\begin{equation}
  \label{eq:54}
 \psi=1+\frac{m_0}{2r},
\end{equation}
where $m_0$ is the Schwarzschild mass.  We have
\begin{align}
  \label{eq:55}
  m &= 2\int_0^\infty \frac{(\psi_{,r})^2}{\psi^2} r^2 \,dr,\\
  &= m_0.
\end{align}
Note that the integral is taken over the two asymptotic regions.

The second example is the Brill-Lindquist\cite{Brill63} initial data.
In this case the data is also time-symmetric and conformally flat, but
here we have $n$ ends $i_k$ and the conformal factor is given by
\begin{equation}
  \label{eq:25}
  \psi= 1+\sum_k^n \frac{m_k}{2r_k},
\end{equation}
where $m_k$ are arbitrary positive constants. The conformal factor
\eqref{eq:25} satisfies \eqref{eq:23} and we have that
\begin{equation}
  \label{eq:29}
 m= \frac{1}{2\pi}\int_{\mathbb{R}^3}\left(\frac{D_a\psi D^a\psi}{\psi^2}
\right) \dvh= \sum_k^n m_k.
\end{equation}

In the non time-symmetric case, we have assumed that the integral of
$K^{ab}K_{ab}\psi^{-8}$ over $\Rt$ is bounded. At infinity, the
integral converges because the assumptions \eqref{eq:39} and
\eqref{eq:6}. At the points $i_k$ the conformal second fundamental
form will, in general, be singular. However the integral will be
bounded because the singular behavior of $K_{ab}$ will be canceled out
by the singular behavior of $\psi$. For example, in the asymptotically
flat case, $K_{ab}=O(r_k^{-4})$ near $i_k$ and then we have that
$K^{ab}K_{ab}\psi^{-8}$ is bounded.  In appendix
\ref{sec:kerr-initial-data} we prove that Kerr initial data satisfy
these conditions.

\section{The variational principle}\label{sec:vari-princ}

In the integral \eqref{eq:114} the mass depends on the metric
variables $\psi$, $q$ (the function $q$ appears in the volume element
and in the indexes contractions) and on the conformal second
fundamental form $K^{ab}$. These functions are not independent, they
have to satisfy the constraint equations \eqref{diver} and
\eqref{Lich}.  In order to formulate the variational principle we want
to express the mass in terms of functions that can be freely varied.
We analyze first the conformal second fundamental form $K^{ab}$ and
the constraint \eqref{diver}.

Consider the following vector field $S^a$
\begin{equation}
  \label{eq:2}
  S_a=K_{ab}\eta^b-\rho^{-2}\eta_a K_{bc}\eta^b\eta^c.
\end{equation}
Using equations \eqref{eq:12}, \eqref{diver} and the Killing equation
for $\eta^a$ it follows that $S^a$ satisfies
\begin{equation}
 \label{eq:J}
\pounds_\eta S^a=0, \quad S^a\eta_a=0, \quad D_a S^a=0.
\end{equation}
From \eqref{eq:52} we deduce an equivalent expression for the total
angular momentum
\begin{equation}
  \label{eq:130}
  J= -\frac{1}{8\pi} \oint_{\Sigma_\infty} S_a  n^a \, \dsh,
\end{equation}
where we have used that the second term in the right-hand side of
\eqref{eq:2} does not contribute to the angular momentum because we
can always chose a closed surface at infinity such that $n^a\eta_a=0$.

The conformal metric $h_{ab}$ can be decomposed into
\begin{equation}
  \label{eq:4}
 h_{ab}=q_{ab}+\rho^{-2} \eta_a \eta_b, 
\end{equation}
where
 \begin{equation}
   \label{eq:5}
   q_{ab} \equiv e^{-2q}(d\rho^2+dz^2),
 \end{equation}
 is the intrinsic metric of the planes orthogonal to $\eta^a$.  Using
 this decomposition and the definition of $S^a$ we obtain the
 following expression for the square of the conformal second
 fundamental form
\begin{equation}
  \label{eq:7}
  K^{ab}K_{ab}=K_{ab}K_{cd}q^{ac}q^{bd} + \rho^{-4}
  (K_{ab}\eta^a\eta^b)^2+ 2\rho^{-2}S^aS_a.
\end{equation}
The two first terms in the right hand side of this equation are
positive, then we have
\begin{equation}
  \label{eq:11}
   K^{ab}K_{ab}\geq 2\rho^{-2}S^aS_a.
\end{equation}
Equations \eqref{eq:130} and \eqref{eq:11} are important because they
show that $S^a$ contains  the angular momentum of $K^{ab}$ and its
square is a lower bound for the square of $K^{ab}$.

We define the tensor
\begin{equation}
  \label{eq:axialpsi}
  \bar K^{ab}=\frac{2}{\eta} S^{(a} \eta^{b)}, 
\end{equation}
we have
\begin{equation}
  \label{eq:17}
 \bar K^{ab}\bar K_{ab}=\frac{2S^aS_a}{\rho^2}. 
\end{equation}
It is interesting to note (but we will not make use of it) that this
tensor is trace free and divergence free. To prove this we use the
Killing equation $D_{(a}\eta_{b)}=0$, the fact that $\eta^a$ is
hypersurface orthogonal, (i.e.; it satisfies $D_a
\eta_b=-\eta_{[a}D_{b]} \ln \eta$) and equations (\ref{eq:J}).

A data will be $(t,\phi)$ symmetric if and only if the following
conditions hold (see \cite{Bardeen70})
\begin{equation}
  \label{eq:ms}
  K_{ab}q^{ac}q^{bd}=0,\quad  K_{ab}\eta^a\eta^b =0.
\end{equation}
This is equivalent to $K^{ab}=\bar K^{ab}$.

The vector $S^a$ can be expressed in terms of a free potential.  Define
the rescaled vector $s^a$ by
\begin{equation}
  \label{eq:13}
  s^a=e^{-2q}S^a,
\end{equation}
then
\begin{equation}
  \label{eq:15}
 \pounds_\eta s^a=0, \quad s^a\eta_a=0, \quad \partial_a s^a=0,
\end{equation}
where $\partial_a$ is the connexion with respect to the flat metric
\begin{equation}
  \label{eq:45}
  \delta =  d\rho^2+dz^2+\rho^2d\varphi^2,
\end{equation}
and in equation \eqref{eq:15} the indexes are moved with this metric
and its inverse. The same will apply to all the equations from now on:
all of them will be given in term of the flat metric $\delta_{ab}$ and
its connexion $\partial_a$.

An arbitrary vector $s^a$, which satisfies equations \eqref{eq:15},
can be written in term of a potential $Y$ in
the following form
\begin{equation}
  \label{eq:axialve}
  s^a=\frac{1}{2\rho^2} \epsilon^{abc} \eta_b  \partial_c Y, 
\end{equation}
where $\epsilon_{abc}$ is the volume element of the flat metric
\eqref{eq:45} and $\pounds_{\eta}Y=0$.  The motivation of the
normalization factor $1/2$ in \eqref{eq:axialve} will be clear in the
next section.  We have the relation
\begin{equation}
  \label{eq:square}
   \bar K^{ab}\bar K_{ab}=\frac{2s^as_a}{\rho^2}=\frac{\partial^aY
   \partial_a Y}{2 \rho^4}.  
\end{equation}
The angular momentum \eqref{eq:J} is given in terms of the potential
$Y$ by
\begin{equation}
  \label{eq:132}
  J=\frac{1}{8}\left (Y(\rho=0,-z) -Y(\rho=0, z)\right ),
\end{equation}
where $z$ is taken to be larger than the location of any point $i_k$.

Motivated by Brill's formula \eqref{eq:114}, we define the mass functional as follows
\begin{equation}
  \label{eq:5c}
 \mf(v,Y)= \frac{1}{32\pi}\int_{\mathbb{R}^3}
  \left(16 \partial_a  v \partial^a v + \rho^{-4} e^{-8v}
    \partial^a Y \partial_a Y
  \right) \dv, 
\end{equation}
where $v = \ln \psi$ and $\dv$ is the flat volume element. Note that
in the integral \eqref{eq:5c} the metric function $q$ does not appear.

From equation \eqref{eq:114} and \eqref{eq:square} we see that for
every axisymmetric and $(t,\phi)$ symmetric data we have $m=\mf(v,Y)$.
From \eqref{eq:11} we see that for every axisymmetric, maximal data,
we have
\begin{equation}
  \label{eq:ineq}
  m\geq\mf(v,Y).
\end{equation}
We emphasize that the functions $(v,Y)$ can be computed for an
arbitrary axisymmetric data (in the construction of the potential $Y$
we have not used the maximal condition) and then the functional
$\mf(v,Y)$ can be also calculated for arbitrary data (provided, of
course, the integral is well defined).  However, only for maximal data
we can use the Brill formula \eqref{eq:114} to conclude
\eqref{eq:ineq} and only for $(t,\phi)$ symmetric data we have that
$\mf(v,Y)$ is in fact the mass.

For the present calculations is more convenient to write the
functional $\mf$ in the form \eqref{eq:5c}, where the axial symmetry
is not explicit.  For completeness, we also write it in a manifest
axisymmetric form
\begin{equation}
\mf(v,Y)= \frac{1}{16}\int^\infty_0 d\rho \int^\infty_{-\infty} dz 
  \left(16\rho \right(v^2_{,z} +v^2_{,\rho}\left) + \rho^{-3} e^{-8v} 
  \right  (Y^2_{,z} +Y^2_{,\rho}\left) 
  \right).  
\end{equation}

Let us define $\af$ as the set of all functions $(v,Y)$ such that the
integral \eqref{eq:5c} is bounded. Although $\mf(v,Y)$ is well defined
in $\af$, not for every function in $\af$ we will have that $\mf(v,Y)$
is equal to the mass of some $(t,\phi)$ symmetric initial data.  This
is a subtle and important point, let us discuss it in detail.  We have
seen that all axisymmetric and $(t,\phi)$ symmetric data can be
generated by three functions $(v,q,Y)$. They are coupled by the
Hamiltonian constraint \eqref{const2}. In coordinates, this equation
is given by
\begin{equation}
  \label{eq:hcq}
  4\frac{\Delta \psi}{\psi} -(q_{,\rho \rho }+q_{,z z })=\frac{\partial^aY
   \partial_a Y}{\rho^4\psi^8},
\end{equation}
where $\Delta$ is the flat Laplacian with respect to \eqref{eq:45}.
For given $(v,Y)$ (remember that $v=\ln\psi$) this is a linear, two
dimensional, Poisson equation for $q$. The delicate point are the
boundary conditions. In order to obtain Brill's formula we have required
that $q$ satisfies \eqref{eq:9} and \eqref{eq:10}. But we cannot impose this two
equations as boundary conditions for a two dimensional Poisson
equation. Let say that we impose \eqref{eq:9} and we ask for solutions
which fall off at infinity. This problem can be solved with an
explicit Green function.  However, in general, the fall off of the
solution will be $q=O(r^{-1})$ which is weaker than \eqref{eq:10}.
Only for some particular source functions $(v,Y)$ the solution $q$
will satisfy \eqref{eq:10}. Let us denote by $\af_1$ the subset of
$\af$ of those functions $(v,Y)$ such that the solution $q$ of
equation \eqref{eq:hcq} satisfies \eqref{eq:9} and \eqref{eq:10}. Only
for functions in $\af_1$ the functional $\mf(v,Y)$ can be written as a
the boundary integral \eqref{eq:111} and hence gives the mass of some
initial data.  A function $v$ of compact support (such that $\psi=1$
near infinity) is an example of a function which is in $\af$ but not
in $\af_1$ (we can take $Y=0$), since in this case clearly $\mf(v,Y)$
is strictly positive and the boundary integral \eqref{eq:111} is zero.

We want to make variations of $\mf(v,Y)$. At first sight, it appears
that the appropriate set for admissible functions is $\af_1$ and not
$\af$.  However, it seems to be difficult to characterize $\af_1$. It
is known how to characterize the set of those $q$ such that
\eqref{eq:hcq} has a solution $\psi$ (for an, essentially, arbitrary
$Y$) which satisfies \eqref{eq:6}, in this case a non-linear equation
must be solved (see \cite{Cantor81b} and \cite{Maxwell05}).  However,
this set is not very useful in the present context since for the Brill
formula is natural to use $(v,Y)$ as independent functions and not
$(q,Y)$.  Instead, what we will do is to take $\af$ as the set of
admissible functions.  Remarkably, it will turn out that the critical
equations in this bigger set are only the stationary, axially
symmetric equations.

Let $\vv$ and $\vY$ be compact supported functions in $\Rt$ with
support in $S$ and such that the support of $\vY$ does not contain the
axis. By equation \eqref{eq:132} we see that this condition implies that
the perturbation $Y+\vY$ does not change the angular momentum of
$Y$. Define
\begin{equation}
  \label{eq:if}
  i(\epsilon)=\mf(v+\epsilon\vv,Y+\epsilon\vY). 
\end{equation}
The first variation of $\mf(v,Y)$ is given by
\begin{equation}
  \label{eq:43}
i'(0)= \frac{1}{16\pi}\int_{\mathbb{R}^3}
  \left(16\partial_a  v \partial^a \vv -4\vv \rho^{-4} e^{-8v}
    \partial^aY\partial_aY + \rho^{-4} e^{-8v} \partial^a Y \partial_a \vY \right) \dv,
\end{equation}
where a prime denotes derivative with respect to $\epsilon$.
Integrating by parts, we obtain that the condition
\begin{equation}
  \label{eq:fvzero}
 i'(0)= 0,  
\end{equation}
for all $\vv$ and $\vY$ is equivalent to the following 
Euler-Lagrange equations
\begin{align}
  \label{eq:elY1}
  4 \Delta v+ \rho^{-4}e^{-8v}\partial^a Y \partial_a Y &=0,\\
  \partial^a(\rho^{-4}e^{-8v}\partial_a Y)=0.\label{eq:elY2}
\end{align}
The second variation is given by
\begin{multline}
  \label{eq:44}
    i''(0)= \frac{1}{16\pi}\int_{\mathbb{R}^3}
  \left\{16\partial_a  \vv \partial^a \vv +\right.\\
\left.\left(32\vv^2      \partial^aY\partial_aY
 - 16\vv  \partial^a Y \partial_a \vY + \partial^a \vY \partial_a \vY
\right)\rho^{-4} e^{-8v}\right\} \dv.  
\end{multline}

There is an equivalent way of deducing equations
\eqref{eq:elY1}--\eqref{eq:elY2}. Instead of taking $Y$ as variable we
 take the vector $s^a$, which should satisfy the constraints
\eqref{eq:15}. The mass functional is given by
\begin{equation}
  \label{eq:5b}
 \mf(v,s)= \frac{1}{8\pi}\int_{\mathbb{R}^3}
  \left(4 \partial_a  v \partial^a v + \rho^{-2} e^{-8v} s^as_a
  \right) \dv.
\end{equation}
Let $\vs^a$ be a compact supported vector in $S$ such that the support
of $\vs^a$ does not contain the axis.  We assume that $\vs^a$
satisfies the constraint
\begin{equation}
  \label{eq:28}
  \partial_a\vs^a=0.
\end{equation}
We define $i$ in analogous way as in \eqref{eq:if}. The first
variation is given by
\begin{equation}
  \label{eq:30}
   i'(0)=  \frac{1}{4\pi}\int_{\mathbb{R}^3}
  \left(4 \partial_a  v \partial^a \vv -4\vv \rho^{-2} e^{-8v} s^as_a+
  \rho^{-2} e^{-8v} s^a\vs_a  
  \right) \dv,
\end{equation}
integrating by parts we get
\begin{equation}
  \label{eq:31}
    i'(0)=   \frac{1}{4\pi}\int_{\mathbb{R}^3}
  \left(-4\vv (\Delta v  + \rho^{-2} e^{-8v} s^as_a)+
  \rho^{-2} e^{-8v} s^a\vs_a  
  \right) \dv.
\end{equation}
From this we deduce the Euler-Lagrange equations
\begin{align}
  \label{eq:22}
  \Delta v+ \rho^{-2}e^{-8v}s^as_a &=0,\\
  \rho^{-2}e^{-8v}s_a=\frac{1}{2}\partial_a \Omega,\label{eq:22b}
\end{align}
for some function $\Omega$. Equation \eqref{eq:22}
follows because we can make arbitrary variations in $\vv$. On the
other hand, variations in $\vs^a$ should satisfy the constraint
\eqref{eq:28}. Writing $\vs^a$ as the curl of an arbitrary vector and
integrating by parts we get
\begin{equation}
  \label{eq:el2}
  \partial_{[a}H_{b]}=0,
\end{equation}
where
\begin{equation}
  \label{eq:H}
  H_a=\rho^{-2}s_ae^{-8v}.
\end{equation} 
Equation \eqref{eq:el2} is equivalent to \eqref{eq:22b}.  Using the
constraint $\partial_as^a=0$, we deduce the following equations which does
not involve $s^a$
\begin{align}
  \label{eq:eO1}
  4 \Delta v+ \rho^{2}e^{8v}\partial^a \Omega\partial_a \Omega &=0,\\
  \partial^a(\rho^{2}e^{8v}\partial_a \Omega)=0.\label{eq:eO2}
\end{align}
Equations \eqref{eq:eO1}--\eqref{eq:eO2} are equivalent to equations
\eqref{eq:elY1}--\eqref{eq:elY2}, the relation between $\Omega $ and
$Y$ is given by
\begin{equation}
  \label{eq:32}
  \partial_a \Omega = \rho^{-4}e^{-8v}\epsilon_{abc}\eta^b\partial^c Y.
\end{equation}
In the next section we will prove that these equations are precisely
the stationary, axisymmetric, vacuum equations.  This will provide
also an interpretation for the potential $Y$ and the velocity $\Omega$
in the stationary case. Note that $Y$ is defined for arbitrary data,
in contrast $\Omega$ is only defined for solutions of the critical
equations, that is, for stationary axisymmetric data.

If we take $Y=0$, then these equations reduce to
\begin{equation}
  \label{eq:weyl}
  \Delta v =0,
\end{equation}
which is Weyl equation for axisymmetric, static, spacetimes. This is of
course consistent with the result that we are going to prove in next
section.  However, it is important to note that the Schwarzschild data
in the form \eqref{eq:54} does not satisfy \eqref{eq:weyl}.
Schwarzschild satisfies \eqref{eq:weyl} in Weyl coordinates where
$\bar v$
and the metric function $\bar q$ are given by
\begin{equation}
  \label{eq:weylco}
\bar v =-\frac{1}{4}\ln\left(\frac{\bar r_+ + \bar r_- -2m}{\bar r_++
    \bar r_- +2m}\right), 
  \quad \bar q = \frac{1}{2}\ln\left(\frac{(\bar r_+ + \bar r_-)^2
      -4m^2}{4\bar r_+\bar r_-}\right),
\end{equation}
with $\bar r^2_{\pm}=\bar \rho^2+(\bar z \pm m)^2$.  The relation
with the isotropic coordinates $(r,\theta)$ used in \eqref{eq:54} is
\begin{equation}
  \label{eq:16}
\bar \rho = \rho \left(1-\frac{m^2}{4r^2} \right),\quad \bar z =
z \left(1+\frac{m^2}{4r^2}\right), 
\end{equation}
where $z=r\cos\theta$ and $ \rho=r\sin\theta$.  Since $X$ is an scalar
independent of coordinates we have $X=\rho^2\psi^4= \bar\rho^2\bar
\psi^4$.  The function $\bar q$ satisfies our assumptions \eqref{eq:9}
and \eqref{eq:10}, however the conformal factor $\bar \psi=e^{\bar v}$
does not satisfies \eqref{eq:23}. The conformal factor is singular on
the rod $\bar \rho=0$, $-m \leq \bar z\leq m$ (which represent the
horizon of Schwarzschild data) and not just on singular points $i_k$.
The integral $\mf(\bar v, 0)$ diverges.  Note that $\Rt$ in Weyl
coordinates $(\bar\rho, \bar z)$ represent the exterior of the black
holes, in contrast to coordinates $(\rho,z)$ where $\Rt$ represent
both asymptotic regions.

\section{Stationary axisymmetric fields}\label{sec:stat-axisymm-fields}

The spacetime metric of a vacuum, stationary and axially symmetric
spacetime can be written, in Weyl coordinates, in the following form
(see, for example, \cite{Wald84})
\begin{equation}
  \label{eq:24}
  g = -V(dt -\sigma d\phi)^2 + V^{-1} \left[\rho^2 d\phi^2 +
    e^{2\gamma}(d\rho^2+dz^2 ) \right],
\end{equation}
where the functions $V$, $\sigma$ and $\gamma$ depend only on
$(\rho,z)$.  The two Killing vectors are
\[
\xi^\mu=\left(\frac{\partial}{\partial t} \right)^\mu, \quad
\eta^\mu=\left(\frac{\partial}{\partial \phi} \right)^\mu,
\] 
they define the scalars
\begin{equation}
  \label{eq:53}
V= - \xi^\mu \xi^\nu g_{\mu\nu} , \quad X = \eta^\mu\eta^\nu
g_{\mu\nu} , \quad
W=\eta^\nu\xi^\mu g_{\nu\mu},   
\end{equation}
where $\mu,\nu$ are spacetime indexes. We have the following relations
\begin{equation}
  \label{eq:33}
  W=V \sigma, \quad \rho^2=V X + W^2.
\end{equation}
The vacuum field equations are given by
\begin{align}
  \partial^a\left (V^{-1}\partial_aV+\rho^{-2}V^2 \sigma
    \partial_a \sigma \right)&=0, \label{eq:stat1}\\
  \partial^a \left(\rho^{-2}V^2\partial_a \sigma \right)&=0.
  \label{eq:stat2}\ 
\end{align}
We want to prove that these equations are equivalent to equations
\eqref{eq:eO1}--\eqref{eq:eO2}. We first compute the relation between
$(V, \sigma)$ and $(v, \Omega)$.

Take an slice $t=constant$ of the metric \eqref{eq:24}. The intrinsic
metric of this surface is given by
\begin{equation}
  \label{eq:34}
  \tilde h = V^{-1} (\rho^2- \sigma^2 V^2)d\phi^2 +
   V^{-1} e^{2\gamma}(d\rho^2+dz^2 ).
\end{equation}
To write this metric in to the form \eqref{eq:1}--\eqref{eq:105} set
\begin{equation}
  \label{eq:35}
  \psi^4=\frac{(\rho^2-V^2\sigma^2)}{V\rho^2}=\frac{X}{\rho^2}=
  \frac{X}{(V X+ W^2)},  
\end{equation}
and
\begin{equation}
  \label{eq:36}
  e^{2q}=\frac{e^{2\gamma}\rho^2}{(\rho^2-V^2\sigma^2)}=\frac{e^{2\gamma}\rho^2}{V X}.     
\end{equation}
From \eqref{eq:35} we deduce
\begin{equation}
  \label{eq:27}
  v(V,\sigma)=\frac{1}{4}\ln{\frac{(\rho^2-V^2\sigma^2)}{V \rho^2}}.
\end{equation}

In order to compute $\Omega(v,\sigma)$ we need to calculate the second
fundamental form of this foliation.  The lapse and the shift of the
foliation $t=constant$ are given by
\begin{equation}
  \label{eq:37}
  N=\frac{\rho}{\sqrt{X}}=\psi^{-2}, \quad   N_a=W(d\phi)_a.
\end{equation}
and the second fundamental form is
\begin{equation}
  \label{eq:38}
  \tilde K_{ab}= -\frac{1}{2N}\tilde D_{(a} N_{b)}.
\end{equation}
We write $N_a$ in terms of the Killing vector $\eta^a$, as in the
previous section we define $\tilde \eta_a= \tilde h_{ab}\eta^b$ where
$\tilde h_{ab}$ is given by \eqref{eq:34}, then we have
\begin{equation}
  \label{eq:40}
 \tilde  \eta_a=\frac{(\rho^2-V^2\sigma^2)}{V}(d\phi)_a.
\end{equation}
Using this expression we write $N_a$ as
\begin{equation}
  \label{eq:41}
  N_a=\Omega\tilde \eta_a,
\end{equation}
where $\Omega$ is given by
\begin{equation}
  \label{eq:26}
  \Omega(V,\sigma)=\frac{V^2\sigma }{2(\rho^2-V^2\sigma^2)}=\frac{W}{X}.
\end{equation}
The scalar $\Omega$ can be interpreted as the angular velocity of the
locally non rotating observers (see \cite{Bardeen70} and also
\cite{Wald84} p. 187). We want to prove that this function $\Omega$ is
precisely the potential $\Omega$ of the previous section. In order to
see this let us compute the vector $s^a$
\begin{equation}
  \label{eq:42}
  s_a=\eta^bK_{ab}=\psi^2\eta^b\tilde K_{ab}=\frac{1}{2N}\psi^2 X
  \partial_a \Omega = \frac{1}{2}\psi^8\rho^2\partial_a\Omega. 
\end{equation}
Where we have used $s_a=s^b\delta_{ab}=S^b h_{ab}$. Equation
\eqref{eq:42} is identical to equation \eqref{eq:22b}. 

Using the relations \eqref{eq:27} and \eqref{eq:26}, after a long but
straightforward computation, we conclude that equations
\eqref{eq:stat1} and \eqref{eq:stat2} for the functions $(V,\sigma)$
are equivalent to equations \eqref{eq:22}--\eqref{eq:22b} for
$(v,\Omega)$.

There is another way to prove the equivalence with the stationary
equations, using the potential $Y$. We replace $v$ by $X$, that is we
consider $X,Y$ as variables.  From equation \eqref{eq:35} we get
\begin{equation}
  \label{eq:46}
  v=\frac{1}{4}\ln X -\frac{1}{2}\ln \rho. 
\end{equation}
Take the functional $\mf$ defined in \eqref{eq:5c} but let us perform
the integral on a bounded domain $B$, in terms of the variables $X,Y$ we get
\begin{equation}
  \label{eq:47}
 \mf(X,Y)= \mf'(X,Y) - \frac{1}{8\pi}\int_{B}
 \ln \left(\frac{\rho}{X}\right )\Delta\ln \rho \dv + \oint_{\partial B}
 \rho^{-1} \ln\left(\frac{\rho}{X}\right) n^a \partial_a\rho \ds,
\end{equation}
where we have defined
\begin{equation}
  \label{eq:49}
  \mf'(X,Y)= \frac{1}{32\pi}\int_{B}
  \left(\frac{ \partial_a  X \partial^a X + \partial^a Y \partial_a Y}{X^2}
  \right) \dv.
\end{equation}
But we have
\begin{equation}
  \label{eq:48}
  \Delta\ln \rho =0,
\end{equation}
for $\rho\neq 0$. Then $\mf$ and $\mf'$ differ only by a boundary
term. Hence they give the same Euler-Lagrange equations.  Note,
however, that the boundary term is singular at the axis $\rho=0$: 
if we take a cylinder $\rho=constant$ near the axis we have, 
$X=O(\rho^2)$, $ds=\rho\, dzd\phi$, $n^a\partial_a\rho=1$, then the
boundary term diverges like  $O(\ln\rho)$ as $\rho\to 0$.   

In \cite{Carter73} Carter formulates a variational principle for the
axisymmetric, stationary equations. This formulation is, essentially,
a modification of the \cite{Ernst68} formulation in which the norm of
the axial Killing vector (and not of the stationary one) is taken to
be the principal variable. 

Carter's Lagrangian is precisely $\mf'$ (we use the same notation for
$X$ and $Y$, this is the reason for the normalization factor $1/2$ in
\eqref{eq:axialve}).  In \cite{Carter73}  it is proved that the
critical equations of $\mf'$ are the stationary, axisymmetric
equations. Therefore, the same is valid for $\mf$.  There are,
however, some important points that we want to stress.

If we ignore boundary terms, then equation \eqref{eq:47} provides an
interpretation for Carter Lagrangian. Also, it gives an interpretation
of the space of admissible functions in which the variations are made
for the following reason.  In Carter's formulation $Y$ is defined in
terms of $W$ and $X$ by
\begin{equation}
  \label{eq:50}
 \epsilon_{abc} \eta^b \partial^c Y = X\partial_a W- W\partial_a X.  
\end{equation}
This equation can easily be obtained from  \eqref{eq:32},
\eqref{eq:35} and \eqref{eq:26}.  
That is, $Y$ is defined only for stationary axisymmetric
spacetimes. From the discussion of section \ref{sec:vari-princ} we
have seen that $Y$ can be defined for arbitrary, axisymmetric
data, and  the variation of $Y$ and $X$ are in fact variation
among axisymmetric and $(t,\phi)$ symmetric data. 

Let us consider boundary terms. The behavior of $X$ near the axis
implies that $\mf'$ is singular if the domain of integration includes
the axis. On the other hand we have seen that $\mf$ is finite.  In
particular, in appendix \ref{sec:kerr-initial-data} we have
explicitly checked that Kerr initial data in quasi-isotropic
coordinates satisfy all our assumptions and then $\mf$ is finite and
equal to the mass for Kerr.  However, is important to note that the
relevant domains of integration  are different in Carter's formulation
and in the present one. In
\cite{Carter73}, the domain is the black hole exterior region, in
which the inner boundary is the horizon. In section
\ref{sec:vari-princ} we have not included any inner boundary
conditions, the domain of integration is the whole manifold which can
include many asymptotic ends. This difference is reflected in the
choice of the coordinate system. 
We have discussed this with Schwarzschild data in section
\ref{sec:vari-princ}. The same apply to 
non-extreme Kerr initial data in Weyl coordinates:  $\mf$ is singular
in this coordinates. However, for extreme Kerr, the Weyl coordinates
and the quasi-isotropic coordinates coincides. In this case both
domains of integration coincides and $\mf$ is finite whether $\mf'$ is
not.

\section{Final comments}\label{sec:final-comments}

We have analyzed the first variation of the, positive definite, mass
functional $\mf$ (defined by \eqref{eq:5c}) over axisymmetric and
$(t,\phi)$ symmetric initial data with fixed angular momentum.  We
have shown that the critical points are the stationary, axial
symmetric equations. This proves the variational principle (ii). The
functional is a lower bound for the mass (inequality \eqref{eq:ineq})
for all maximal, axisymmetric data.  This proves (i'). In order to
prove (i), and hence inequality \eqref{eq:14}, we should prove that
extreme Kerr is the unique absolute minimum of  $\mf$ over axisymmetric and
$(t,\phi)$ symmetric initial data with fixed angular momentum. This
will require the study of the second variation of
$\mf$, given in equation \eqref{eq:44}.  

\section*{Acknowledgments}
It is a pleasure to thank Abhay Ashtekar, Marc Mars and Walter Simon
for  valuable discussions. 

This work has been supported by the Sonderforschungsbereich SFB/TR7 of
the Deutsche Forschungsgemeinschaft.

\appendix
\section{Kerr initial data}\label{sec:kerr-initial-data}
Consider the Kerr metric in Boyer-Lindquist coordinates $(t,\tilde
r,\theta,\phi)$. The scalars \eqref{eq:53} are given by
\begin{align}
  \label{eq:56}
  V &=\frac{\Delta-a^2\sin^2\theta}{\Sigma}, \quad W= -\frac{2ma\tilde
    r\sin^2\theta}{\Sigma},\\
  \label{eq:57}
  X & =\left(\frac{(\tilde r^2+a^2)^2 -\Delta a^2
      \sin^2\theta}{\Sigma} \right)\sin^2\theta,
\end{align}
where
\begin{equation}
  \label{eq:58}
\Delta =\tilde r^2+a^2-2m\tilde r, \quad 
\Sigma=\tilde r^2+a^2 \cos^2 \theta,
\end{equation}
and $m$ is the total mass and $a$ is the angular momentum per unit
mass (i.e. $J=ma$).

The intrinsic metric $\tilde h_{ab}$ of a hypersurface $t=constant$ in
these coordinates is given by
\begin{equation} 
\label{tildehBL}
\tilde h =\frac{\Sigma}{\Delta} d\tilde r^2+ \Sigma d\theta^2 +\eta 
d \phi^2.
\end{equation}
The metric \eqref{tildehBL} has a coordinate singularity when
$\Delta=0$. The solutions of the equation $\Delta =0$ are given by
\begin{equation} \label{r+-}
\tilde r_+=m+\sqrt{m^2-a^2}, \quad \tilde r_-=m-\sqrt{m^2-a^2}.
\end{equation}
By the following coordinate transformation we extend the metric to a
complete manifold with two asymptotic ends. 
Let us define the quasi-isotropic radius $r$ as the positive root of
the following equation
\begin{equation}
\tilde r = r +m+\frac{m^2-a^2}{4 r}.
\end{equation}
Note that when $a=0$ this reduce to the isotropic radius for the
Schwarzschild metric. The manifold (like in the Schwarzschild case) has
to isometric asymptotically flat components (the region $\tilde r \geq
\tilde r_+$ of the metric \eqref{tildehBL}) joined at the minimal
surface (the horizon) $\tilde r= \tilde r_+$. The components of
$\tilde h_{ab}$ in the coordinates $( r , \theta, \phi)$ are given by
\begin{equation}
  \label{eq:tildehi}
\tilde h  =\frac{\Sigma}{ r^2 } d  r^2+ \Sigma d\theta^2 +\eta 
d \phi^2.  
\end{equation}
The metric \eqref{eq:tildehi} has the form
\eqref{eq:1}--\eqref{eq:105} with
\begin{equation}
  \label{eq:51}
  \psi^4= \frac{X}{\rho^2}, \quad
  e^{-2q}=\frac{\sin^2\theta\Sigma}{X}, 
\end{equation}
where $ \rho= r\sin\theta$ and $ z= r\cos\theta$.  Assume $m> |a|$.
Then, from \eqref{eq:51} we see that in the limit $r\to 0$ we have
\begin{equation}
  \label{eq:59}
 \psi = \frac{\sqrt{m^2-a^2}}{ r}
 +\frac{m}{2\sqrt{m^2-a^2}}+O( r) ,\quad  \psi_{, r}= O(
 r^{-2}), 
\end{equation}
and at infinity
\begin{equation}
  \label{eq:61}
 \psi =1+\frac{m}{2 r}+ O(r^{-2}),\quad q=O(r^{-2}). 
\end{equation} 
From \eqref{eq:51} we also have that   
\begin{equation}
  \label{eq:60}
   q(\rho=0)=0.
\end{equation}
Hence, $q$ satisfies \eqref{eq:9}, \eqref{eq:10} and $\psi$ satisfies
\eqref{eq:6} and \eqref{eq:23}.
 
The velocity $\Omega$ can be calculated from equation
\eqref{eq:26} using \eqref{eq:56} and \eqref{eq:57}
\begin{equation}
  \label{eq:19}
 \Omega=- \frac{2ma\tilde r}{(\tilde r^2+a^2)^2 -\Delta a^2 \sin^2\theta}.
\end{equation}

The potential $Y$ is given by
\begin{equation}
  \label{eq:18}
Y = 2ma(\cos^3\theta-3\cos\theta)- \frac{2ma^3\cos\theta\sin^4\theta}{\Sigma}. 
\end{equation}
Note that equation \eqref{eq:132} is satisfied for $z \neq 0$. To see
that the integral of $\partial^aY\partial_aY \rho^{-4} \psi^{-8}$ over
$\Rt$ is bounded we need to check the behavior of this function at
infinity and at the axis $\rho=0$. 
At infinity we have
\begin{equation}
  \label{eq:21}
\partial^aY\partial_aY \rho^{-4}= O(r^{-6}), 
\end{equation}
and at the axis
\begin{equation}
  \label{eq:62}
\partial^aY\partial_aY \rho^{-4}\psi^{-8}=O(r^2),
\end{equation}
where we have used the \eqref{eq:59}. Then, the integral is bounded
and therefore we have proved that the Kerr initial data satisfies our
assumptions which implies that $\mf(v,Y)=m$.

Weyl coordinates $(\bar\rho, \bar z)$ are related to the coordinates
$(r,\theta)$ by
 \begin{equation}
   \label{eq:64}
  \bar \rho= \sqrt{\Delta} \sin\theta, \quad \bar z = (\tilde r
  -m)\cos\theta.   
 \end{equation}
 
Consider now the extreme case $m=|a|$. In this case we have
\begin{equation}
  \label{eq:65}
  r=\tilde r -m,\quad \Delta = r^2,
\end{equation}
and the coordinates $(r,\theta)$ are equal to the Weyl coordinates.  
Equations \eqref{eq:60} and \eqref{eq:61} are still valid in this
case. The fall off of the conformal factor near $r=0$ is however different
\begin{equation}
  \label{eq:66}
 \psi = \frac{\sqrt{2m}}{(1+\cos^2\theta)^{1/4} \sqrt{r}} +
 O(r^{1/2}) ,\quad  \psi_{, r}= O(r^{-3/2}),   
\end{equation}
this is because $r=0$ is not an asymptotically flat in this
case. Nevertheless $\psi$ satisfies   \eqref{eq:23}. 
The fall of behavior of $Y$ at infinity is the same as in the
non-extreme case. Near the axis, because of \eqref{eq:66},  we have
\begin{equation}
  \label{eq:63}
\partial^aY\partial_aY \rho^{-4}\psi^{-8}=O(r^{-2}),
\end{equation}
and hence we conclude that $\partial^aY\partial_aY \rho^{-4}\psi^{-8}$
is integrable over $\Rt$. 


\end{document}